\documentclass[twocolumn,pre,aps,showpacs,a4paper,floatfix,amssymb]{revtex4-1}

\usepackage{psfrag}
\usepackage{pstricks}
\usepackage{graphicx}
\graphicspath{{figures/}}
\usepackage{subfigure}
\usepackage{hyperref}
\usepackage{color}
\usepackage{amsmath}
\usepackage{soul}

\hypersetup{
  colorlinks,
  citecolor=[rgb]{0,0,1},
  linkcolor=[rgb]{0,0,1},
  urlcolor=[rgb]{0,0,1}}

\newcommand{\kBT}{\,k_\mathrm{B}T}
\newcommand{\refeq}[1]{Eq.~(\ref{#1})}
\newcommand{\refref}[1]{Ref.~\cite{#1}}
\newcommand{\refig}[1]{Fig.~\ref{#1}}
\newcommand{\HGPFQ}{{}_pF_q}
\newcommand{\ud}{\mathrm{d}}

\newcommand{\MAi}[1]{#1} 
\begin{document}

\date{\today}
\title{Dissociation rates from single-molecule pulling experiments under large thermal fluctuations or large applied force}
\author{Masoud Abkenar$^{1,2}$, Thomas H. Gray$^{1,3}$, Alessio Zaccone$^{1,3}$}
\affiliation{${}^1$Statistical Physics Group, Department of Chemical
Engineering and Biotechnology, University of Cambridge, CB2 3RA Cambridge, U.K.}
\affiliation{${}^2$Lehrstuhl f\"ur Biophysik, Physik-Department, Technische Universit\"at M\"unchen, 85748 Garching, Germany} 
\affiliation{${}^3$Cavendish Laboratory, University of Cambridge, CB3 0HE Cambridge,
U.K.}

\begin{abstract}
Theories that are used to extract energy-landscape information from single-molecule pulling experiments in biophysics are all invariably based on Kramers' theory of thermally-activated escape rate from a potential well. As is well known, this theory recovers the Arrhenius dependence of the rate on the barrier energy, and crucially relies on the assumption that the barrier energy is much larger than $k_{B}T$ (limit of comparatively low thermal fluctuations). As was already shown in Dudko, Hummer, Szabo Phys.\ Rev.\ Lett.\ (2006), this approach leads to the unphysical prediction of dissociation time increasing with decreasing binding energy when the latter is lowered to values comparable to $k_{B}T$ (limit of large thermal fluctuations). We propose a new theoretical framework (fully supported by numerical simulations) which amends Kramers' theory in this limit, and use it to extract the dissociation rate from single-molecule experiments where now predictions are physically meaningful and in agreement with simulations over the whole range of applied forces (binding energies). These results are expected to be relevant for a large number of experimental settings in single-molecule biophysics. 
\end{abstract}

\maketitle

\section{Introduction}
Formation of intermolecular bonds in thermally agitated environments is ubiquitous from biological to condensed matter physics.
Examples include nanoparticle adsorption to membranes \cite{wilhelm2003intracellular,dasgupta1402}, 
protein-ligand bindings \cite{merkel1999energy,friddle2012interpreting,brujic2006single}, mechanical studies of cellular membranes \cite{evans1995sensitive,ramms2013keratins} and conformational changes of proteins \cite{yu2015}
The energy gain upon forming the bond - the \emph{binding energy} $Q$ - 
determines the bond's stability in an environment with thermal fluctuations.
When the binding energy and thermal fluctuations (energies of the order of $\kBT$) are comparable in size, i.e. $E\equiv Q/\kBT \approx 1$, the bond can break on a relatively short time scale. 
Conversely, larger binding energies mean increased stability and longer mean dissociation times.

\begin{figure}
\includegraphics[width=1\columnwidth]{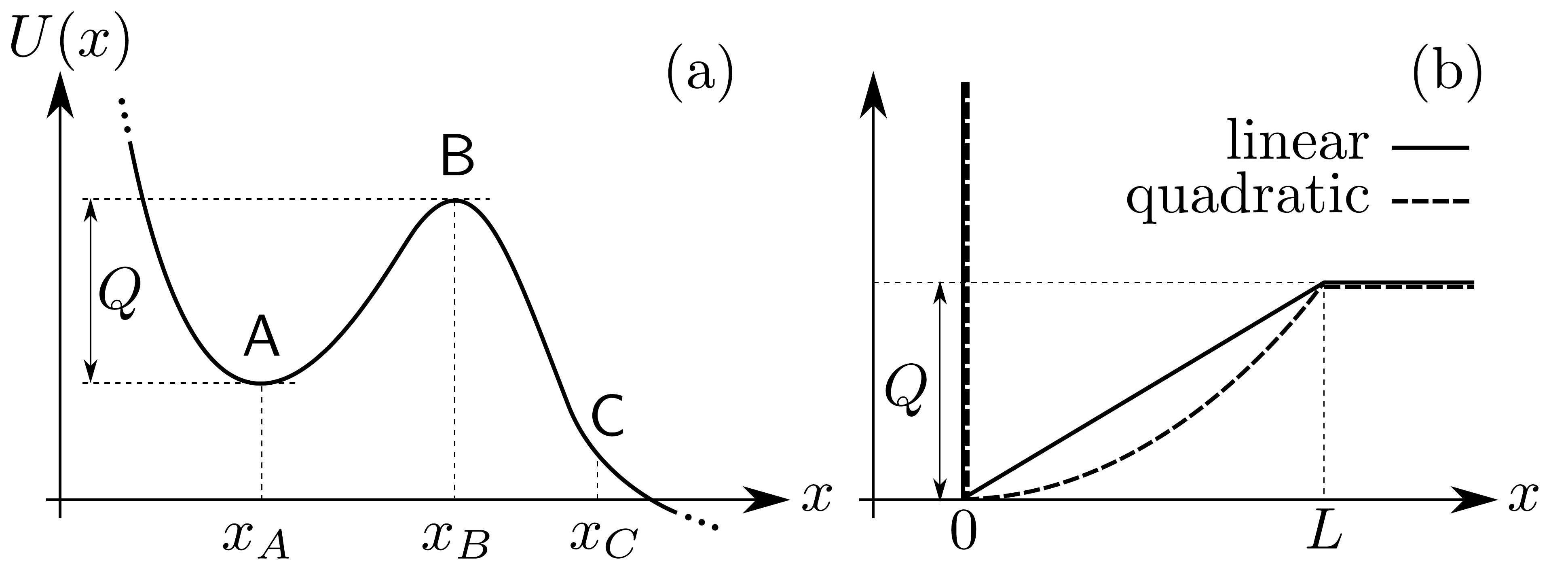}
\caption{
(a) Schematic of the potential used in the derivation of Kramers' dissociation time, \refeq{eq:kramers}. 
\MAi{(b)} The truncated linear [\refeq{eq:ramppot}] and quadratic [\refeq{eq:quadratic}] potentials used in the analytic calculation of dissociation times.
}
\label{fig:kramers}
\end{figure}

Bond stability is described by the mean dissociation time $\tau$ or, alternatively, by the dissociation rate $r=1/\tau$.
Kramers \cite{kramers1940brownian} recovered the Arrhenius law \cite{arrhenius1889reaktionsgeschwindigkeit} ($\tau \propto \text{ exp}(E)$) from the Smoluchowski equation and the energy landscape displayed in \refig{fig:kramers}(a):
\begin{equation}
\tau=\frac{2\pi\kBT}{(|\omega_A| |\omega_B|)^{1/2}D} \exp\left(\frac{Q}{\kBT}\right),
\label{eq:kramers}
\end{equation}
where 
$\omega_{A,B}=\partial^2 U(x)/\partial x^2 |_{x=x_A,x_B}$ are the potential's curvatures at its minimum  and maximum respectively, and $D$ is the diffusion constant.

In order to reach the above result Kramers had to make a series of assumptions. One of the most important is that a stationary energy distribution exists around the minimum. This imposes the requirement of large energy barriers \cite{kramers1940brownian,hanggi1990reaction}. Kramers used the saddle-point approximation to evaluate two integrals. By doing so, he imposed implicitly the condition of steepness in the energy landscape around the minimum and maximum. Evidently these assumptions fail for small values of $E$, which corresponds both to shallow potential wells and large thermal energies. Furthermore, some potentials (e.g. Lennard-Jones (LJ)) do not have an energy barrier over which escape takes place. In these cases, it is clear that \refeq{eq:kramers} cannot be applied, but Arrhenius-law-style behaviour of $\tau(E)$ appears to exist for sufficiently deep wells.

We use an alternative mean first-passage time formalism - the Ornstein-Uhlenbeck (OU) method - to obtain solutions $\tau(E)$ for situations where Kramers' assumptions break down: dissociation of dimers interacting through a LJ potential and single-molecule constant-force pulling experiments. First, analytic results are found for truncated linear and quadratic potential wells and used as crude approximations to the (non-analytic) LJ and linear-cubic potentials. Secondly, numerical solutions for these non-analytic potentials are obtained.

Our numerical model gives results that agree excellently with data from dissociation simulations of LJ dimers (see \refig{fig:dtime_vs_E}). Surprisingly good agreement between our analytic models and the data is observed also. The data for larger values of $E$ appear to have an Arrhenius-style $E$-dependence. However, significant deviation from this exponential behaviour is found for smaller values of $E$, with 27 and 13-fold discrepancies for $E=0$ and 1 respectively. Our approach can be applied successfully to constant-force pulling experiments and, importantly, for small $E$ our model does not give an unphysical divergence in $\tau$, which is the case for the model based on Kramers' theory presented in \refref{dudko2006intrinsic}.

For a particle moving in a potential energy landscape $U(x)$ with reflecting/absorbing boundaries at $x_A$/$x_B$ respectively, and an initial position $x$ satisfying $x_A \leq x \leq x_B$, the OU method gives the mean first-passage time, up to a multiplicative constant, as \cite{uhlenbeck1930theory,gardiner2004handbook}:

\begin{equation}
\tau = C \int_{x}^{x_B} \ud y \exp\left(\frac{U(y)}{\kBT}\right)\int_{x_A}^y \ud z \exp\left( -\frac{U(z)}{\kBT}\right),
\label{eq:OU}
\end{equation}


\subsection{Truncated linear potential}
\noindent First, the truncated linear potential [solid line in \refig{fig:kramers}(b)]:
\begin{equation}
U(x) = 
\begin{cases} 
\frac{Qx}{L} & 0 < x \leq L\\
Q & x > L \\
\end{cases}.
\label{eq:ramppot}
\end{equation}
 Placing the reflecting and absorbing boundaries at $x_A=0$ and $x_B=L$ produces:
\begin{eqnarray}
\tau(x)&=& \frac{L^2}{D}\frac{1}{(Q/\kBT)^2}  \left[ \exp\left(\frac{Q}{\kBT} \right) - \exp\left(\frac{Qx}{L\kBT} \right) \right] \nonumber \\ 
&+& \frac{L\left(x-L\right)}{D}\frac{1}{Q/\kBT}.
\end{eqnarray}
Setting $x=0$ gives $\tau(0)$: the mean first-passage time for a particle starting at the reflecting boundary. This is the quantity of interest because the linear (and quadratic) wells are used as crude models of the LJ landscape describing the interactions of Brownian dimers. The dimers start in the potential minimum and so we start at the point lowest in potential also, which corresponds to $x=0$ - the reflecting wall. Multiplying by $D/L^2$ renders the above dimensionless: $\tau_0 = D\tau(0)/L^2$:
\begin{eqnarray}
\tau_0=\frac{1}{\left(Q/k_BT\right)^2} \left[ \exp\left(\frac{Q}{\kBT} \right) - 1 \right]- \frac{1}{Q/\kBT}.
\nonumber
\end{eqnarray}

\noindent Finally, substituting $E=Q/\kBT$, we have
\begin{eqnarray}
\tau_0=\frac{1}{E^2} [\exp (E) - 1] - \frac{1}{E},
\label{eq:ramp}
\end{eqnarray}
with a sub-exponential (non-Arrhenius) dependence ($\sim \exp (E)/E^2$) for larger values of $E$. 
Note that as $E \rightarrow 0$, \refeq{eq:ramp} remains finite and the free diffusion limit is recovered for $E=0$
\cite{footnoteLimit}.

Next, the truncated quadratic well:
\begin{equation}
U(x) = 
\begin{cases} 
\frac{Qx^2}{L^2} & 0 < x \leq L\\
Q & x > L. \\
\end{cases}
\label{eq:quadraticpot}
\end{equation}
Proceeding as before we find:
\begin{eqnarray}
\tau\MAi{(x)} &=& \frac{L^2}{2D}\left[{}_2 F_2\left[\{1,1\},\left\{\frac{3}{2},2\right\},E\right]\right] \\
&-& x^2\frac{L^2}{2D}\left[{}_2 F_2\left[\{1,1\},\left\{\frac{3}{2},2\right\},\left(\frac{x}{L}\right)^2 E \right]\right], \nonumber
\label{eq:quadratic}
\end{eqnarray}
where $\HGPFQ(a_{1},...,a_{p};b_{1},...,b_{q};z)$ is the generalized hypergeometric function \cite{HGPFQ}. 
We form $\tau_0$:
\begin{equation}
\tau_0 = \frac{1}{2}{}_2 F_2\left[\{1,1\}, \left\{\frac{3}{2},2\right\}, E\right].
\label{eq:quadratic1}
\end{equation}

The hypergeometric function notation obscures the behaviour of $\tau$ making immediate comparison with the result for the linear well difficult. However, we can deduce the asymptotic behaviours. As $E \rightarrow 0$ the force due to the potential also falls towards zero. This means that the escape process transforms into free diffusion and thus the results must coincide with each other, and the free-diffusion limit, for $E=0$.

\subsection{Force-tilted potentials}
The OU method can be used for one-dimensional potential energy landscapes only. Justification of its applicability to LJ dimers and the single-molecule pulling experiment (see later) is necessary. First, the dimers: the LJ potential is spherically symmetric and thus a function of the \textit{absolute distance} between the particles - a \textit{scalar} quantity - alone. The escape process is thus effectively one-dimensional. Secondly, the pulling experiment: applying a pulling-force defines a preferential direction for escape. We exploit this by characterising the unfolding process as motion in an energy landscape defined by one coordinate along the line of the force, and so reduce the dimensionality from three to one.

\begin{figure}[b]
\subfigure[]{\includegraphics[width=0.45\columnwidth]{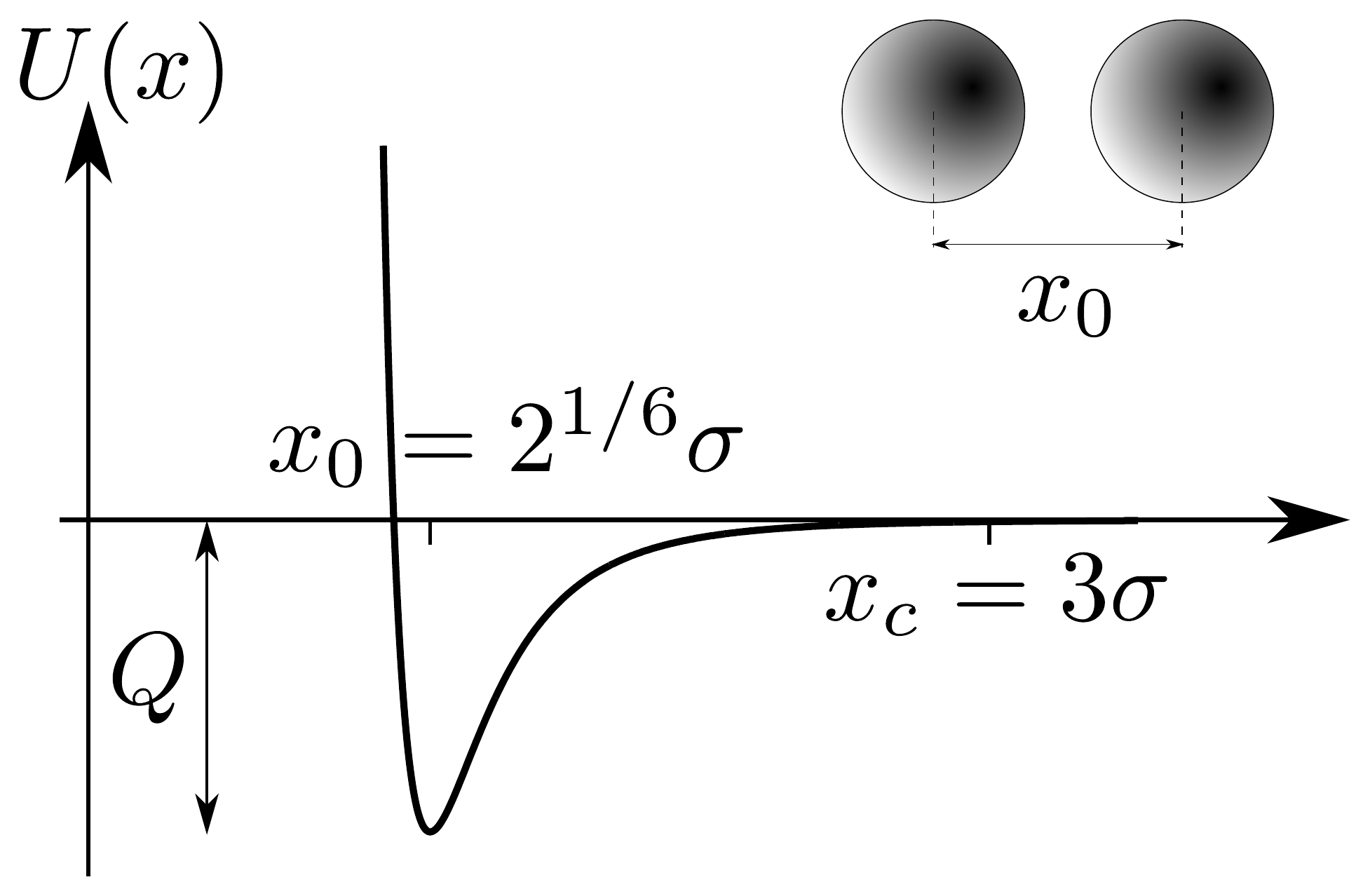}\label{fig:LJ}}
\subfigure[]{\includegraphics[width=0.53\columnwidth]{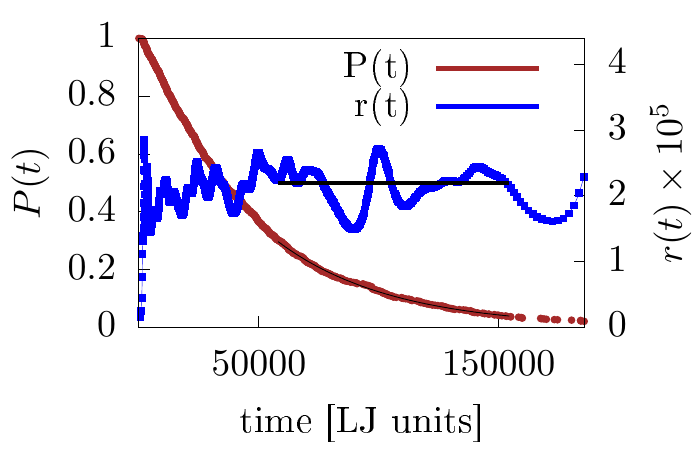}\label{fig:Pt}}

\caption{{\bf Simulation setup and analysis.} 
(a) The binary system and interaction potential used in simulations.
For each value of $E$ between $500$ to $1000$ pairs were simulated.
(b) The association ratio $P(t)$ (brown circles) 
and the dissociation rate $r(t)$ (blue squares)
vs time for $E=5.0$.
}\label{fig:simulation} 
\end{figure}

In order to test the analytical theory, we conducted Brownian Dynamics (BD) simulations of two initially bonded Brownian dimers.
The particles interact through a Lennard-Jones (LJ) potential, 
$U(x)=4 Q [ ({\sigma}/{x})^{12} - ({\sigma}/{x})^{6} ]$.
The potential is truncated at $x=3\,\sigma$ and shifted by $U(3\,\sigma)$ to avoid a discontinuity at $x=3\,\sigma$.
Here, $x$ is the distance between the particles, $Q$ is the depth of the potential well, 
and $\sigma$ is the characteristic particle size.
The particles are initialized at a distance of $x_0=2^{1/6}\sigma$ from each other, 
where $U(x)$ has its minimum.
The simulations are done in the framework of Brownian dynamics, 
i.e., the overdamped Langevin equation with a negligible inertial term,
using the LAMMPS molecular dynamics package \cite{plimpton1995fast}.
The friction coefficient in the overdamped Langevin equation is $\gamma=10^3$ in LJ units, $(mQ/\sigma^2)^{1/2}$. We measure the dissociation time $\tau$ for different values of $E$ according to the following protocol. 

Different values of $E$ were simulated by fixing $Q=1$ and changing the temperature $T$ from $0.1$ to $10$, where $T$ is measured in LJ units, $Q/k_\mathrm{B}$.
Depending on the temperature, different time steps were used for the time integration, from $\delta t=0.001$ (low temperatures) to $\delta t=0.0001$ (high temperatures), 
where time is measured in LJ units: $(m\sigma^2/Q)^{1/2}$.
$500$ to $1000$ dimers were simulated independently for each value of $E$.
The simulations were run for a number of steps $N$ in the range $N=10^7$ to $1.5\times 10^9$,
depending on $E$. $N$ was chosen such that for each $E$, a substantial number of dimers dissociated before the simulation finished at time $t_\mathrm{end}=N \delta t$.
$\tau$ is defined as the first time at which the two particles are separated by a distance greater than $3\,\sigma$, see \refig{fig:Pt}. This dissociation condition was implemented in the OU method by placing the absorbing barrier at $x_B=3\sigma$.

The mean dissociation times are calculated using the ``survival'' function $P(t)$, which measures the fraction of bonds that are still intact at time $t$. \refig{fig:Pt} shows $P(t)$ for a sample simulation run.
The instantaneous dissociation rate is defined as 
$r(t) = -{d\ln P(t)}/{dt}$ \cite{boilley1993nuclear} 
which is also plotted (blue curve) in \refig{fig:Pt}. 
The steady-state dissociation rate $r$ is calculated from the plateau of $r(t)$ \cite{footnotePoisson}.
\MAi{Since we change $E$ by changing the temperature $T$, 
each simulation point has a different diffusion constant $D=\kBT/\gamma$.
Therefore, to compare simulation results with our models, 
we use $\tau=T/r$ from the simulations,
so that the scaling factor $D/L^2$ (used to make our models dimensionless) is compensated for.}

\section{Results}
%


Attempting to fit the Arrhenius law ($\tau \propto \text{exp}(E)$) to the data for large values of $E$ allows us to test whether or not Kramers' method retains some applicability despite the requirement of an energy barrier to escape no longer being fulfilled.

\begin{figure}
\includegraphics[width=1.01\columnwidth]{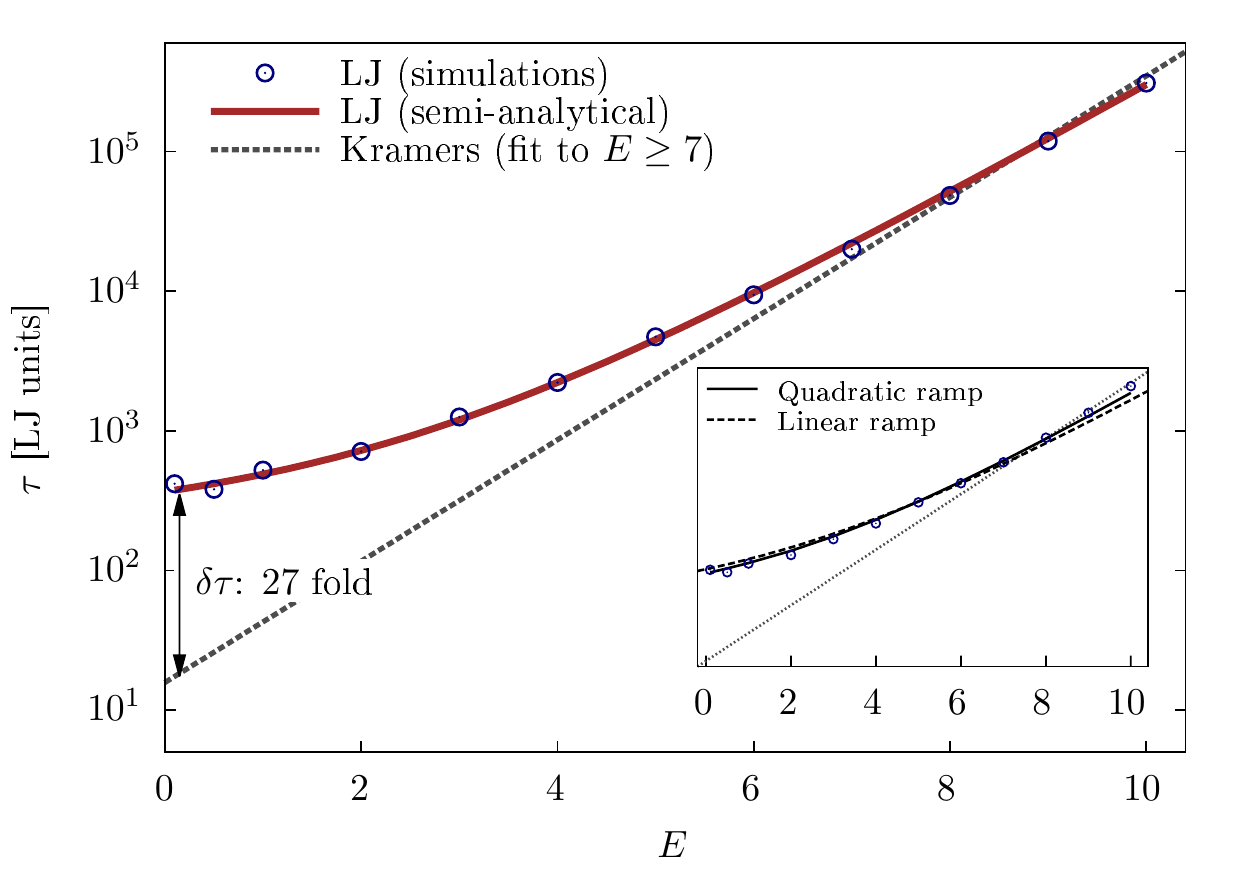}
\caption{{\bf Dissociation time vs bond strength for a binary system.} 
Kramers' law works well for $E\gtrsim7$, 
while the analytical solution, and quadratic potential model, presented here fit better across the entire range.
Only one adjustable prefactor $C$ that shifts the curves in the $y$ direction is used to fit the theoretical models to the data.
Corrections to Kramers' law can be as large as $27$-fold in the dissociation time. The inset shows the comparison between the same simulation data and the analytical solutions of \refeq{eq:OU} with the simplified quadratic and linear ramp potentials. 
}
\label{fig:dtime_vs_E}
\end{figure} 

\refig{fig:dtime_vs_E} shows the simulation data and the result obtained from the OU method (numerical integration of \refeq{eq:OU}) with the LJ potential energy landscape: excellent agreement is observed. The inset of \refig{fig:dtime_vs_E} shows the same data but with the analytic results found previously for the linear (\refeq{eq:ramp}) and quadratic (\refeq{eq:quadratic1}) potential wells. Given the stark differences between the linear and LJ energy landscapes, it is interesting to see good agreement between the linear model (\refeq{eq:ramp}) and the data. Perhaps this hints that the salient features of an energy landscape - the width and depth of the well - play a significant part in determining the broad first-passage properties, with the finer features producing smaller corrections. Better agreement still is observed between the quadratic model and the simulation data. This is because the quadratic potential can capture some of the LJ potential's curvature, which is not possible for the linear potential.

The plots also show the Arrhenius law fitted to the data for $E \gtrsim 7$. This range was chosen in the hope that Kramers' requirement of a relatively deep and steep well was met. Should this be the case, then we might conclude that deviations from the Arrhenius law for larger values of $E$ are attributable to the shape of the well alone. Unsurprisingly, the Arrhenius law deviates significantly from the simulation data for smaller values of $E$ showing that, as expected, it is unable to predict correctly dissociation times for weak bonds/large thermal fluctuations. More data is required before we can be sure of a deviation from the Arrhenius law for large values of $E$, as predicted by our analytic models and the numerical integration.

Moreover, the analytic solution of \refeq{eq:quadratic1} agrees well with the simulations even without adjustable parameters.
For example, the best fit to the data using \refeq{eq:ramp}, gives $L^2/D=763$.
The corresponding prefactor in the simulation is $\gamma/w^2$,
where $\gamma=10^3$ is the friction coefficient and $w$ is the effective length associated with the Lennard-Jones interaction.
Given $\sigma=1$ used in the simulations, we expect $1< w<2$ based on the analytical prediction.
Equating $C= \gamma/w^2$ gives $w=1.4$, which agrees well with the theoretical expectation.

A method based on Kramers' theory for extracting the mean rupture rates from single-molecule constant-force pulling experiments is presented in Ref.~\cite{dudko2006intrinsic}. Comparison to Brownian dynamics simulations of a system with a cubic potential-energy landscape showed discrepancies for large pulling forces, where the energy barrier to rupture is comparable in size to the thermal fluctuations Ref.~\cite{dudko2006intrinsic}.

\begin{figure}
\includegraphics[width=1.01\columnwidth]{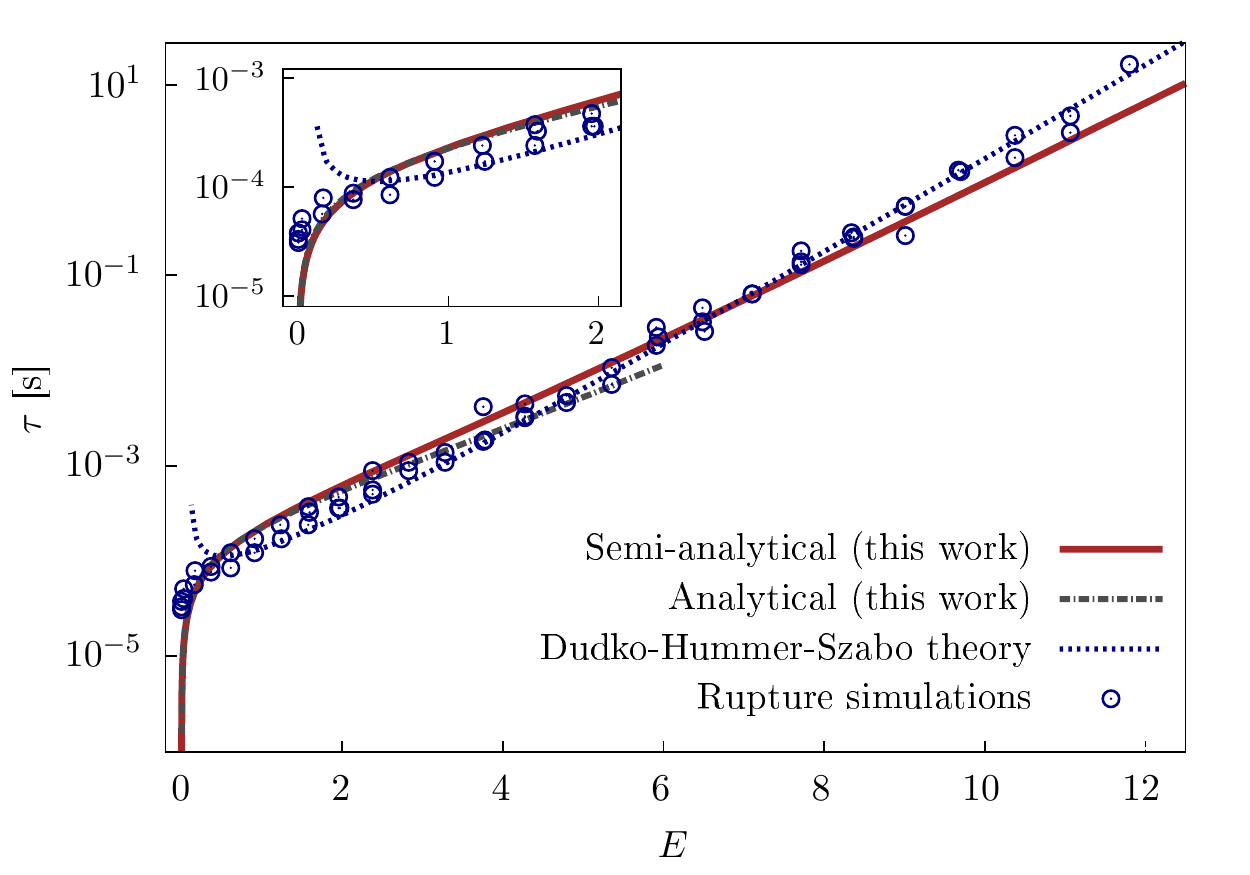}
\caption{{\bf Dissociation time in simulated pulling experiments.} 
Dissociation time $\tau$ vs effective bond-strength $E$ calculated from constant-force simulated pulling experiments. The potential used is the same cubic potential of Ref.~\cite{dudko2006intrinsic}, with an attractive minimum and an energy barrier, qualitatively similar to \refig{fig:kramers}(a). 
While the Kramers-Dudko-Hummer-Szabo theory \cite{dudko2006intrinsic} breaks down for $E\lesssim 1$ (large pulling force $F$),
the full integration of the OU theory \refeq{eq:OU} captures the decrease in $\tau$ for very low $E$, which highlights the importance of the effect. The inset is a zoom-in of the low-$E$ regime. 
}
\label{fig:dudko}
\end{figure}

We use the OU method with the model potential given in Ref.~\cite{dudko2006intrinsic} to calculate mean rupture times and compare with the aforementioned simulations. Both numerical integration of the full potential and a new analytical expression (a linearised approximation) valid for large forces $F$ (small $E$) are used. The analytical expression and its derivation can be found in the Appendix.
Applying a force $F$ alters the energy landscape by lowering the barrier to rupture and moving the minimum and maximum closer together. Thus, new $x_A$, $x_B$, and $Q$ must be calculated for each value of $F$. We consider forces ranging in size from zero up to the value at which the barrier height vanishes and the minimum and maximum overlap.
Both methods yield, up to a multiplicative constant, the corresponding rupture times. 
This constant is determined by fitting to the simulation data.
Results are plotted in \refig{fig:dudko}, as a function of the \textit{effective} binding energy in units of $k_\text{B}T$, 
together with the model and simulation data from Ref.~\cite{dudko2006intrinsic}.


For $1 \lesssim E \lesssim 12$, corresponding to $F<280\mathrm{pN}$,
both the Dudko-Kramers theory and our numerical integration match the simulation data well. For larger forces - an important regime for experimental single-molecule studies - the Dudko-Kramers theory breaks down, as already discussed in \cite{dudko2006intrinsic}, predicting an unphysical result: 
the mean rupture time increasing upon decreasing the effective binding energy.
Instead, both our analytical solution and numerical calculations predict the correct decrease in $\tau$ with decreasing $E$, in excellent agreement with the simulation data.
Also, both our approaches recover the vanishing rupture time in the limit $E\rightarrow 0$ \cite{footnoteIdentically}.
%


\section{Conclusion}
Kramers' theory for the escape time from a potential well has been extremely successful in providing a theoretical foundation to the Arrhenius law in physics, chemistry and biology. 
However, the underlying assumptions restrict its applicability to deep wells with a barrier to escape due  and its validity for shallow wells/large thermal fluctuations has not been properly investigated. This limit is crucial for biophysics: in single-molecule pulling experiments, an external force is applied with a cantilever to explore the energy landscape of proteins and to determine the dissociation time of receptor-ligand complexes. Clearly, the effective well depth in this case is controlled by the applied force and can become comparable to, if not smaller than, the thermal fluctuations.
Kramers' theory breaks down dramatically in this limit, but our approach, making use of the Ornstein-Uhlenbeck method, produces models that have been verified against numerical simulations.

We then applied this approach to a single-molecule pulling experiment and found that our models provide an excellent description of simulation data even in the large-force limit, where the famous approach of Dudko, Hummer and Szabo, based on Kramers' theory, provides unphysical results.
Our method can be applied widely and, in particular, it may play a major role in the quantitative analysis of force-spectroscopy experiments in biological systems.

\begin{acknowledgements}
We are very grateful to E. M. Terentjev and M. Rief for many useful discussions and input.
\end{acknowledgements}

\appendix
\section{Analytical approximation for the rate of single-molecule pulling experiments}
We will demonstrate how the Ornstein-Uhlenbeck (OU) method can be used to analyse the data presented by Dudko, Hummer and Szabo in \cite{dudko2006intrinsic}. In doing so, we will convert the effect of an applied force $F$ into the lowering of a potential barrier, which allows the OU method to be applied and mean first-passage times  $\tau$ to be obtained. Two cases will be considered: numerical integration of the full linear-cubic potential for all $F$; and analytic integration of a linearised version of the potential valid for larger forces, corresponding to lower barriers.

The potential energy landscape under consideration is a linear-cubic combination:
\begin{equation}
U_0(x) = \frac{3\Delta G^\ddagger}{2} \frac{x}{x^\ddagger} - 2\Delta G^\ddagger \left(\frac{x}{x^\ddagger}\right)^3
\end{equation}

\noindent A biasing force is then applied, which leads to the following potential defined in terms of the effective quantities $x_c^\ddagger$ and $\Delta G_c^\ddagger$ -  the \textit{apparent} minimum-to-barrier distance and \textit{apparent} barrier height:

\begin{equation}
U(x) = \frac{3\Delta G_c^\ddagger}{2} \frac{x}{x_c^\ddagger} - 2\Delta G_c^\ddagger \left(\frac{x}{x_c^\ddagger}\right)^3 - Fx
\end{equation}
where $F$ is the applied biasing force. \par

\par
\par \noindent In order to apply the OU method we need two quantities: the form of the potential $U(x)$, which we know, and the coordinates of the minimum and maximum of the potential well. However, in order to compare the mean first-passage times produced by this method to data from the paper we also need to know the barrier height $Q$.  The stationary points are found in the next section and the barrier height determined in the third.

\subsection{The stationary points}
As usual, these are found by solving the equation $\frac{\text{d}U}{\text{d}x} = 0$.
\begin{equation}
U(x) = \left(\frac{3\Delta G_c^\ddagger}{2} - Fx_c^\ddagger \right)\frac{x}{x_c^\ddagger} - 2\Delta G_c^\ddagger \left(\frac{x}{x_c^\ddagger}\right)^3
\end{equation}

\begin{equation}
\frac{\text{d}U}{\text{d}x} = \left(\frac{3\Delta G_c^\ddagger}{2} - Fx_c^\ddagger \right) - 6\Delta G_c^\ddagger \left(\frac{x}{x_c^\ddagger}\right)^2 = 0
\end{equation}

\begin{equation}
\Rightarrow \frac{x_\pm}{x_c^\ddagger} = \pm \sqrt{\frac{1}{6\Delta G_c^\ddagger}\left(\frac{3\Delta G_c^\ddagger}{2} - Fx_c^\ddagger \right)}
\end{equation}

\subsection{The barrier height}
In the previous section we saw that $x_+$ - the positive root of $\frac{\text{d}U}{\text{d}x} = 0$ - is exactly minus the negative root. Combining this with the antisymmetry of $U(x)$ means that $\tilde{Q}$ is given by: $\tilde{Q} = U(x_+) - U(x_-) = 2U(x_+)$:

\begin{equation}
\tilde{Q} = 2 \times \left(\frac{3\Delta G_c^\ddagger}{2} - Fx_c^\ddagger \right) \frac{x_+}{x_c^\ddagger} - 2 \times 2\Delta G_c^\ddagger \left(\frac{x_+}{x_c^\ddagger}\right)^3
\end{equation}

\begin{eqnarray}
\tilde{Q} &=& 2\left(\frac{3\Delta G_c^\ddagger}{2} - Fx_c^\ddagger \right)\left(\frac{3\Delta G_c^\ddagger}{2} - Fx_c^\ddagger \right)^{1/2} \left(\frac{1}{6\Delta G_c^\ddagger}\right)^{1/2} \nonumber \\
&-& 4\Delta G_c^\ddagger \left(\frac{1}{6\Delta G_c^\ddagger}\right)\left(\frac{1}{6\Delta G_c^\ddagger}\right)^{1/2}\left(\frac{3\Delta G_c^\ddagger}{2} - Fx_c^\ddagger \right)^{3/2}
\end{eqnarray}

\begin{equation}
\tilde{Q} = \frac{4}{3}\left(\frac{1}{6\Delta G_c^\ddagger}\right)^{1/2} \left(\frac{3\Delta G_c^\ddagger}{2} - Fx_c^\ddagger \right)^{3/2}
\end{equation}

\subsection{Applying the Ornstein-Uhlebeck method}
Using the result presented by Gardiner in \cite{gardiner2004handbook}, the mean first-passage time $\tau$ is given up to a multiplicative constant by the following formula which appears out the front of the expression and shall be called $C$:

\begin{eqnarray}
\tau = C\int_{x_-}^{x_+} \text{d}y \text{ exp}\left(\frac{U(y)}{k_\text{B}T}\right) \int_{x_-}^y \text{d}z \text{ exp}\left(-\frac{U(z)}{k_\text{B}T}\right)
\end{eqnarray}

In Ref.~\cite{dudko2006intrinsic}, we are told that $\Delta G_c^\ddagger = 17.6 k_\text{B}T$. Plugging in this information to the above formula leads to the following:

\begin{eqnarray}
\tau &=& C\int_{x_-}^{x_+} \text{d}y \text{ exp}\left[(26.4-Fx_c^\ddagger)\frac{y}{x_c^\ddagger} - 35.2\left(\frac{y}{x_c^\ddagger}\right)^3\right] \nonumber \\
&\times& \int_{x_-}^y \text{d}z \text{ exp}\left[35.2\left(\frac{z}{x_c^\ddagger}\right)^3 - (26.4-Fx_c^\ddagger)\frac{z}{x_c^\ddagger}\right]
\end{eqnarray}

\noindent We see that the only quantities in the above expression remaining to be evaluated are $x_\pm$ and $Fx_c^\ddagger$. From equation (5) we notice that evaluating $x_\pm$ boils down to determining $Fx_c^\ddagger$. The values of $F$ can be read off the graph presented in \cite{dudko2006intrinsic} and $x_c^\ddagger$ is given there too as 0.34nm. From here nothing more is required apart from the conversion of $Fx_c^\ddagger$ from Joules to thermal units, $k_\text{B}T$. To do this, we assumed a standard temperature of 298K.

\subsection{Linearisation of the potential for large thermal fluctuations (small barriers)}
For sufficiently large forces, the potential energy landscape between the minimum (starting point) and maximum (exit point) appears almost linear. This hints at the option of modelling the landscape as a linear potential between these two points which will allow us to obtain an analytic form valid for large forces/small barriers (low $Q/k_\text{B}T$). In the rest of this section we will work through how this is achieved and ultimately provide the final analytic result.

The linear ramp potential will run from $(x_-, U(x_-))$ to $(x_+, U(x_+))$. The symmetry properties of the potential mean that $x_+ = -x_-$ and $U(x_+) = - U(x_-)$. Thus the gradient of the line connecting these two points is $m = \frac{U(x_+)}{x_+}$ and hence the linear potential is:
\begin{equation}
V(x) = U(x_-)+ \frac{U(x_+)}{x_+}(x-x_-)
\end{equation}

\begin{figure}[h]
\includegraphics[width = 1\columnwidth]{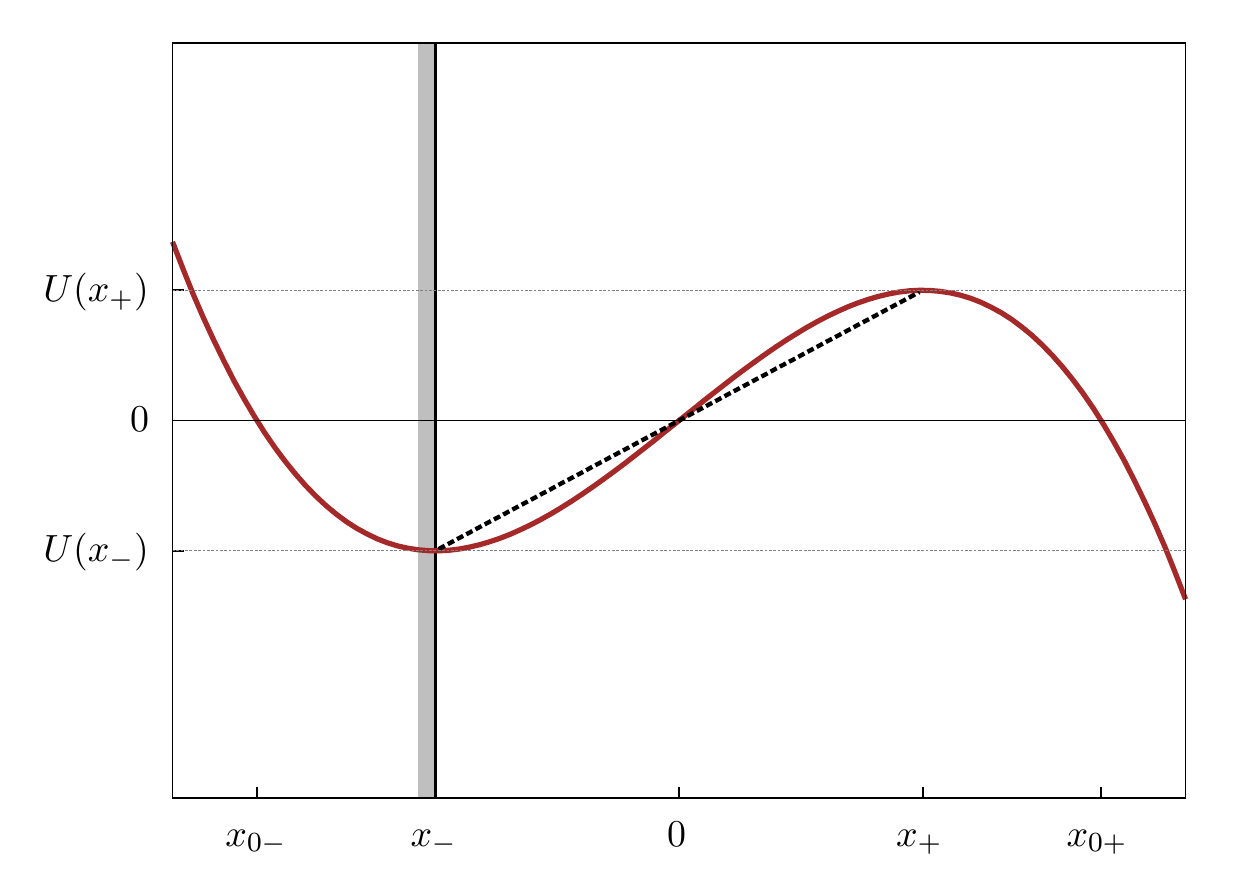}
\caption{A schematic depicting the forms of the linear-cubic potential and the linear approximation to this potential. The linear approximation works well when the applied force is large, corresponding to small energy barriers $Q/k_\text{B}T$. The vertical line at $x_-$ represents the reflecting wall boundary condition applied at this point.}
\label{fig:linearised}
\end{figure}

\subsection{Performing the Integration}
The OU method provides the mean first-passage time for a general starting point $x$ in the region $(x_-,x_+)$ as:

\begin{eqnarray}
\tau(x) &=& C\int_{x}^{x_+} \text{d}y \text{ exp}\left(\frac{V(y)}{k_\text{B}T}\right) \nonumber \\ &\times& \int_{x_-}^y \text{d}z \text{ exp}\left(-\frac{V(z)}{k_\text{B}T}\right)
\end{eqnarray}

\noindent Inserting the expression for $V$ into the above and evaluating the integrals leads to the following expression for $\tau(x)$:

\begin{eqnarray}
\frac{\tau(x)}{C} &=& \left(\frac{k_\text{B}Tx_+}{U(x_+)}\right)^2 \text{exp}\left(\frac{2U(x_+)}{k_\text{B}T}\right)\nonumber \\ &-& \left(\frac{k_\text{B}Tx_+}{U(x_+)}\right)^2\text{exp}\left(\frac{U(x_+)}{k_\text{B}Tx_+}(x-x_-)\right) \nonumber \\ &-& \frac{k_\text{B}Tx_+}{U(x_+)} (x_+-x)
\end{eqnarray}

\subsection{Specialising to the case of Dudko-Hummer-Szabo model simulations}
We now set the initial position $x$ in the above to the position of the minimum in the linear-cubic potential - $x_-$. This gives the following:

\begin{eqnarray}
\tau(x_-) &=& C\left(\frac{k_\text{B}T}{U(x_+)}\right)^2 x^2_+ \left[\text{exp}\left(\frac{2U(x_+)}{k_\text{B}T}\right) - 1\right] \nonumber \\ &-& 2C\frac{k_\text{B}T}{U(x_+)}x_+^2
\end{eqnarray}

\subsection{Evaluating $\tau(x_-)$}
In order to evaluate $\tau(x_-)$ we need to put in the expressions for $x_+$ and $U(x_+)$. These are as follows:

\begin{equation}
x_+ = x_c\left(\frac{1}{6\Delta G_c^\ddagger}\right)^{1/2}\left(\frac{3\Delta G_c^\ddagger}{2} - Fx_c^\ddagger\right)^{1/2}
\end{equation}

\begin{equation}
U(x_+) = \frac{2}{3}\left(\frac{1}{6\Delta G_c^\ddagger}\right)^{1/2}\left(\frac{3\Delta G_c^\ddagger}{2} - Fx_c^\ddagger\right)^{3/2}
\end{equation}

\par
\noindent We may now determine the following quantities of use:

\begin{equation}
\frac{U(x_+)}{k_\text{B}T} = \frac{2}{3}\left(\frac{1}{4} - \frac{Fx_c^\ddagger}{6\Delta G_c^\ddagger}\right)^{1/2}\left(\frac{3\Delta G_c^\ddagger}{2k_\text{B}T} - \frac{Fx_c^\ddagger}{k_\text{B}T}\right)
\end{equation}

\begin{equation}
\frac{k_\text{B}Tx_+}{U(x_+)} = \frac{3x_c}{2}\left(\frac{3\Delta G_c^\ddagger}{2k_\text{B}T} - \frac{Fx_c^\ddagger}{k_\text{B}T}\right)^{-1}
\end{equation}

\begin{equation}
\frac{k_\text{B}Tx_+^2}{U(x_+)} = \frac{3x_c^2}{2}\left(\frac{1}{6\Delta G_c^\ddagger/k_\text{B}T}\right)^{1/2}\left(\frac{3\Delta G_c^\ddagger}{2k_\text{B}T} - \frac{Fx_c^\ddagger}{k_\text{B}T}\right)^{-1/2}
\end{equation}

\subsection{Final result}
Plugging in all of the above expressions into the formula for $\tau(x_-)$ gives the following result:

\begin{eqnarray}
\frac{\tau(x_-)}{Cx_c^2} &=& \text{exp}\left[\frac{4}{3}\left(\frac{3\Delta G_c^\ddagger}{2k_\text{B}T} - \frac{Fx_c^\ddagger}{k_\text{B}T}\right)\left(\frac{1}{4} - \frac{Fx_c^\ddagger}{6\Delta G_c^\ddagger}\right)^{1/2}\right] \nonumber \\ &\times& \frac{9}{4}\left(\frac{3\Delta G_c^\ddagger}{2k_\text{B}T} - \frac{Fx_c^\ddagger}{k_\text{B}T}\right)^{-2} - \frac{9}{4}\left(\frac{3\Delta G_c^\ddagger}{2k_\text{B}T} - \frac{Fx_c^\ddagger}{k_\text{B}T}\right)^{-2} \nonumber
\\ &-& 3\left(\frac{1}{6\Delta G_c^\ddagger/k_\text{B}T}\right)^{1/2}\left(\frac{3\Delta G_c^\ddagger}{2k_\text{B}T} - \frac{Fx_c^\ddagger}{k_\text{B}T}\right)^{-1/2}
\end{eqnarray}

\noindent For a given value of the applied force $F$, the barrier height $Q$ is given by:
\begin{equation}
\frac{Q}{k_\text{B}T} = \frac{4}{3}\left(\frac{1}{4} - \frac{Fx_c^\ddagger}{6\Delta G_c^\ddagger}\right)^{1/2}\left(\frac{3\Delta G_c^\ddagger}{2k_\text{B}T} - \frac{Fx_c^\ddagger}{k_\text{B}T}\right)
\end{equation}

\noindent and plotting $\tau(x_-)/Cx_c^2$ vs $Q/k_\text{B}T$ enables comparison with the results from the numerical integration of the full cubic-linear potential and simulation data from the Dudko paper.

\subsection{Comparison}
From the graph of ``Rate" vs ``Force" in \cite{dudko2006intrinsic}, we can obtain the mean first-passage time by taking 1/``Rate". The integrals in Eq.(10) were evaluated numerically for different values of the applied force $F$, and mean first-passage times were obtained up to the multiplicative constant $C$. Comparing these sets of data allowed the scaling factor $C$ to be determined and hence the model fitted to the data, see Fig.4 in the main article.

For the linearised potential, the analytic result was evaluated for a range of values of $F$. Again, this produced a series of mean first-passage times up to a multiplicative constant $C$. These quantities were scaled to fit the data in an identical fashion to that described above (also for this, see Fig.4 in the main article).

A plot of the numerically integrated result and the analytic result (from linearisation) shows excellent agreement with the data for low barrier heights $Q/k_\text{B}T$ (large applied forces $F$) as shown in Fig.4 in the main article.


\begin{thebibliography}{21}
\bibitem{footnoteLimit}
\MAi{\refeq{eq:ramp} in the limit of $E\rightarrow 0$ yields $\tau_0 = 1/2 + E^3/6 +\mathcal{O}(E^4)$ which recovers the 1D free diffusion limit $\tau_0=1/2$.}
\bibitem{footnotePoisson}
\MAi{Since the dissociation is a homogeneous Poisson process, the rate can be} alternatively derived from fitting a shifted exponential function $P_0 \exp[-r(t-t_0)]$ to the $P(t)$ curve. The two methods give indistinguishable results.
\bibitem{footnoteIdentically}
Here the force is so strong that minimum and maximum points coincide and the mean first-passage time is identically zero.
\end{thebibliography}

\end{document}